\documentclass[11pt]{article}

\usepackage{fullpage}
\usepackage{latexsym}

\def\01{\{0,1\}}
\def\eps{\varepsilon}
\def\R{\mbox{\bf R}}
\def\OR{\mbox{\rm OR}}
\def\MAJ{\mbox{\rm MAJ}}
\def\PARITY{\mbox{\rm PARITY}}
\def\Pr{\mbox{\rm Pr}}
\newcommand{\ket}[1]{|#1\rangle}

\newcommand{\unif}{\mbox{\scriptsize\it unif}}

\newtheorem{definition}{Definition}[section]
\newtheorem{theorem}[definition]{Theorem}

\newtheorem{lemma}[definition]{Lemma}
\newtheorem{corollary}[definition]{Corollary}

\newenvironment{proof}
{\noindent {\bf Proof. }}
{{\hfill $\Box$}\\
\smallskip
}

\begin{document}

\title{Average-Case Quantum Query Complexity\thanks{A preliminary version of
this paper appeared in the Proceedings of 17th Annual Symposium on Theoretical Aspects of Computer Science (STACS'2000), Springer, LNCS 1770, 2000.}}
\author{
Andris Ambainis\thanks{Part of this work was done when visiting Microsoft Research. Supported by Microsoft Research Fellowship and NSF Grant CCR-9800024.}\\
Computer Science Department\\
University of California\\
Berkeley CA 94720\\
USA\\
{\tt ambainis@cs.berkeley.edu}
\and
Ronald de Wolf\thanks{Partially supported by the EU fifth framework
project QAIP, IST--1999--11234. Also affiliated with the ILLC, University of Amsterdam.}\\
CWI\\ 
P.O.~Box 94079\\
1090 GB Amsterdam\\
The Netherlands\\
{\tt rdewolf@cwi.nl}
}

\maketitle

\begin{abstract}
We compare classical and quantum query complexities of total Boolean functions. 
It is known that for {\em worst-case} complexity, the gap between 
quantum and classical can be at most polynomial~\cite{bbcmw:polynomials}.
We show that for {\em average-case} complexity under the uniform distribution,
quantum algorithms can be exponentially faster than classical algorithms.
Under non-uniform distributions the gap can even be super-exponential.
We also prove some general bounds for average-case complexity
and show that the average-case quantum complexity of MAJORITY under the uniform
distribution is nearly quadratically better than the classical complexity.
\end{abstract}

\section{Introduction}

The field of quantum computation studies the power of computers based on
quantum mechanical principles.
So far, most quantum algorithms---and {\em all} physically 
implemented ones---have operated in the so-called {\em black-box} setting.
In the black-box model, the input of the function $f$ that we want 
to compute can only be accessed by means of queries to a ``black-box''.
This returns the $i$th bit of the input when queried on $i$. 
The complexity of computing $f$ is measured by the required number 
of queries. In this setting we want quantum algorithms that
use significantly fewer queries than the best classical algorithms.
Examples of quantum black-box algorithms that are provably better 
than any classical algorithm can be found in
\cite{deutsch&jozsa,simon:power,grover:search,bht:collision,bhmt:countingj,betal:distinctness}.
Even Shor's quantum algorithm for period-finding, which is the core of his efficient 
factoring algorithm~\cite{shor:factoring}, can be viewed as a black-box 
algorithm~\cite{cleve:orderfinding}. 

We restrict our attention to computing 
total Boolean functions $f$ on $N$ variables.
The query complexity of $f$ depends on the kind of errors one allows. 
For example, we can distinguish between exact computation, zero-error 
computation (a.k.a.~Las Vegas), and bounded-error computation (Monte Carlo).
In each of these models, {\em worst-case} complexity is usually considered:
the complexity is the number of queries required for the ``hardest'' input.
Let $D(f)$, $R(f)$ and $Q(f)$ denote the worst-case query complexity of
computing $f$ for classical deterministic algorithms, classical randomized
bounded-error algorithms, and quantum bounded-error algorithms, respectively.
More precise definitions will be given in the next section.
Since quantum bounded-error algorithms are at least as powerful 
as classical bounded-error algorithms, and classical bounded-error algorithms are at least
as powerful as deterministic algorithms, we have $Q(f)\leq R(f)\leq D(f)$.
The main quantum success here is Grover's algorithm~\cite{grover:search}.
It can compute the \OR-function with bounded-error using $\Theta(\sqrt{N})$
queries (which is optimal~\cite{bbbv:str&weak,bbht:bounds,zalka:grover}).
Thus $Q(\OR)\in\Theta(\sqrt{N})$, whereas $D(\OR)=N$ and $R(\OR)\in\Theta(N)$.
This is the biggest gap known between quantum and classical worst-case
complexities for total functions.
(In contrast, for {\em partial} Boolean functions the gap can be
much bigger~\cite{deutsch&jozsa,simon:power,cleve:orderfinding}.)
In fact, it is known that the gap between $D(f)$ and $Q(f)$ is at most polynomial
for {\em every} total $f$: $D(f)\in O(Q(f)^6)$~\cite{bbcmw:polynomials}.
This is similar to the best known relation between classical deterministic and
randomized algorithms: $D(f)\in O(R(f)^3)$~\cite{nisan:pram&dt}.

Given some probability distribution $\mu$ on the set of inputs $\01^N$ one may
also consider {\em average-case} complexity instead of worst-case complexity.
Average-case complexity concerns the {\em expected} number of queries needed
when the input is distributed according to $\mu$.
If the hard inputs receive little $\mu$-probability, then average-case 
complexity can be significantly smaller than worst-case complexity.
Let $D^\mu(f)$, $R^\mu(f)$, and $Q^\mu(f)$ denote the average-case
analogues of $D(f)$, $R(f)$, and $Q(f)$, respectively,
to be defined more precisely in the next section.
Again $Q^\mu(f)\leq R^\mu(f)\leq D^\mu(f)$.
The objective of this paper is to compare these measures and to 
investigate the possible gaps between them.
Our main results are:
\begin{itemize}
\item Under uniform $\mu$, $Q^\mu(f)$ and $R^\mu(f)$ can be 
super-exponentially smaller than $D^\mu(f)$.
\item Under uniform $\mu$, $Q^\mu(f)$ can be exponentially smaller 
than $R^\mu(f)$.
Thus the polynomial relation that holds between quantum and classical 
query complexities in the case of worst-case complexity~\cite{bbcmw:polynomials}
does not carry over to the average-case setting.
\item Under non-uniform $\mu$ the gap can be even larger:
we give distributions $\mu$ where $Q^\mu(\OR)$ is constant,
whereas $R^\mu(\OR)$ is almost $\sqrt{N}$.
\item For every $f$ and $\mu$, $R^\mu(f)$ is lower bounded by the expected 
{\em block sensitivity} 
$E_\mu[bs(f)]$ and $Q^\mu(f)$ is lower bounded by $E_\mu[\sqrt{bs(f)}]$.
\item For the MAJORITY-function under uniform $\mu$, we have that
$Q^\mu(f)\in O(\sqrt{N}(\log N)^2)$ and $Q^\mu(f)\in\Omega(\sqrt{N})$. 
In contrast, $R^\mu(f)\in\Omega(N)$.
\item For the PARITY-function, the gap between $Q^\mu$ and $R^\mu$ can
be quadratic, but not more. 
Under uniform $\mu$, PARITY has $Q^\mu(f)\in\Omega(N)$.
\end{itemize}

\section{Definitions}\label{secdefs}

Let $f:\01^N\rightarrow\01$ be a Boolean function.
This function is {\em symmetric} if $f(X)$ only depends on $|X|$, 
the Hamming weight (the number of 1s) of $X$. 
We will in particular consider the following symmetric functions:
$\OR(X)=1$ iff $|X|\geq 1$; $\MAJ(X)=1$ iff $|X|>N/2$; $\PARITY(X)=1$
iff $|X|$ is odd.
If $X\in\01^N$ is an input and $S$ a set of (indices of) variables,
we use $X^S$ to denote the input obtained by flipping the values of
the $S$-variables in $X$.
The {\em block sensitivity} $bs_X(f)$ of $f$ on an input $X$ is the
maximal number $b$ for which there are $b$ disjoint sets of variables
$S_1,\ldots,S_b$ such that $f(X)\neq f(X^{S_i})$ for all $1\leq i\leq b$.
The block sensitivity $bs(f)$ of $f$ is $\max_X bs_X(f)$.

We are interested in the question how many bits of the input have 
to be queried in order to compute $f$, either for the worst-case or
average-case input. We assume familiarity with classical computation
and briefly sketch the definition of quantum query algorithms.
For a general introduction to quantum computing, see the book of Nielsen 
and Chuang~\cite{nielsen&chuang:qc}. For more details about (quantum) query
complexity we refer to~\cite{buhrman&wolf:dectreesurvey}.

An $m$-qubit state is a $2^m$-dimensional unit vector 
of complex numbers, written $\sum_{x\in\01^m}\alpha_x\ket{x}$.
The complex number $\alpha_x$ is called the {\em amplitude} 
of the basis state $\ket{x}$.
A $T$-query quantum algorithm corresponds to a unitary transformation
$$
A=U_TOU_{T-1}O\ldots U_1OU_0.
$$
Here the $U_j$ are unitary transformations on $m$ qubits.
These $U_j$ are independent of the input.
Each $O$ corresponds to a query to the input $X\in\01^N$, formalized
as the unitary transformation
$$
\ket{i,b,z}\rightarrow\ket{i,b\oplus x_i,z}.
$$
Here $i\in\{1,\ldots,N\}$, $b\in\01$, $\oplus$ is addition modulo 2,
and $z\in\01^{m-\log N-1}$ is the workspace, which remains unaffected by the query.
Intuitively, $O$ just gives us the bit $x_i$ when queried on $i$.
We will sometimes use the word ``oracle'' to refer to $X$ 
as well as to the corresponding $O$.
The initial state of the algorithm is the all-zero state $\ket{0^m}$. 
The final state is $A\ket{0^m}$, which depends on the input $X$ 
via the $T$ queries that are made. A measurement of a dedicated
{\em output bit} of the final state will yield the output.
It can be shown that this linear-algebraic quantum model is at least 
as strong as classical randomized computation: 
any classical $T$-query randomized algorithm can be simulated 
by a $T$-query quantum algorithm having the same error probabilities.

As described above, the quantum algorithm will make exactly $T$ queries
on {\em every} input $X$.
Since we are interested in average-case number of queries and 
the required number of queries will depend on the input $X$,
we need to allow the algorithm to give an output after fewer 
than $T$ queries. We will do that by measuring, after each $U_j$,
a dedicated {\em flag-qubit} of the intermediate state at that point
(this measurement may alter the state).
This bit indicates whether the algorithm is already prepared 
to stop and output a value. If this bit is 1, then we measure the output bit,
output its value $A(X)\in\01$ and stop; if the flag-bit is 0 we let 
the algorithm continue with the next query $O$ and $U_{j+1}$.
Note that the number of queries that the algorithm makes on input $X$ is
now a random variable, since it depends on the probabilistic outcome
of measuring the flag-qubit after each step. 
We use $T_A(X)$ to denote the {\em expected} number of queries 
that $A$ makes on input $X$.
The Boolean output $A(X)$ of the algorithm is a random variable as well.

We mainly focus on three kinds of algorithms for computing $f$: 
classical {\em deterministic}, classical {\em randomized} bounded-error, 
and {\em quantum} bounded-error algorithms.
Let ${\cal D}(f)$ denote the set of classical {\em deterministic} 
algorithms that compute $f$. Let 
${\cal R}(f)=\{\mbox{classical }A\mid\forall X\in\01^N:\Pr[A(X)=f(X)]\geq
2/3\}$ 
be the set of classical {\em randomized} algorithms that compute $f$ 
with bounded error probability. The error probability $1/3$ is
not essential; it can be reduced to any small $\eps$ by running 
the algorithm $O(\log(1/\eps))$ times and outputting 
the majority answer of those runs. Similarly we let 
${\cal Q}(f)=\{\mbox{quantum }A\mid\forall X\in\01^N:\Pr[A(X)=f(X)]\geq
2/3\}$ be the set of bounded-error {\em quantum} algorithms for $f$.
We define the following worst-case complexities:
\begin{eqnarray*}
D(f) & = & \min_{A\in{\cal D}(f)}\max_{X\in\01^N}T_A(X)\\
R(f) & = & \min_{A\in{\cal R}(f)}\max_{X\in\01^N}T_A(X)\\
Q(f) & = & \min_{A\in{\cal Q}(f)}\max_{X\in\01^N}T_A(X)
\end{eqnarray*}
$D(f)$ is also known as the {\em decision tree complexity} of $f$
and $R(f)$ as the {\em bounded-error} decision tree complexity of $f$.
Since quantum computation generalizes randomized computation
and randomized computation generalizes deterministic
computation, we have $Q(f)\leq R(f)\leq D(f)\leq N$ for all $f$.
The three worst-case complexities are polynomially related:
$D(f)\in O(R(f)^3)$~\cite{nisan:pram&dt} and 
$D(f)\in O(Q(f)^6)$~\cite{bbcmw:polynomials} for all total $f$.

Let $\mu:\01^N\rightarrow[0,1]$ be a probability distribution.
We define the {\em average-case complexity} of an algorithm
$A$ with respect to a distribution $\mu$ as:
$$
T_A^\mu=\sum_{X\in\01^N}\mu(X)T_A(X).
$$
The average-case deterministic, randomized, and quantum complexities 
of $f$ with respect to $\mu$ are
\begin{eqnarray*}
D^\mu(f) & = & \min_{A\in{\cal D}(f)}T^\mu_A\\
R^\mu(f) & = & \min_{A\in{\cal R}(f)}T^\mu_A\\
Q^\mu(f) & = & \min_{A\in{\cal Q}(f)}T^\mu_A
\end{eqnarray*}
Note that the algorithms still have to satisfy the appropriate output
requirements (such as outputting $f(X)$ with probability $\geq 2/3$ 
in case of $R^\mu$ or $Q^\mu$) on {\em all} inputs $X$, 
even on $X$ that have $\mu(X)=0$.
Clearly $Q^\mu(f)\leq R^\mu(f)\leq D^\mu(f)\leq N$ for all $\mu$ and $f$.
Our goal is to examine how large the gaps between these measures can be, 
in particular for the uniform distribution $\mbox{\it unif\/}(X)=2^{-N}$.

The above treatment of average-case complexity is the standard one used 
in average-case analysis of algorithms~\cite{vitter&flajolet:av}. 
One counter-intuitive consequence of these definitions, however, is that 
the average-case performance of polynomially related algorithms can be 
superpolynomially apart (we will see this happen in Section~\ref{secnonunifgap}).
This seemingly paradoxical effect makes these definitions unsuitable
for dealing with polynomial-time reducibilities and average-case 
complexity classes, which is what led Levin to his alternative 
definition of ``polynomial time on average''~\cite{levin:av}.%
\footnote{We thank Umesh Vazirani for drawing our attention to this.}
Nevertheless, we feel our definitions are the appropriate ones for 
our query complexity setting: they {\em are} just the average numbers of
queries that one needs when the input is drawn according to distribution $\mu$.

\section{Super-Exponential Gap between $D^{\unif}(f)$ and $Q^{\unif}(f)$}

Before comparing the power of classical and quantum computing, we first
compare the power of {\em deterministic} and {\em bounded-error} algorithms.
It is not hard to show that $D^{\unif}(f)$ can be much larger 
then $R^{\unif}(f)$ and $Q^{\unif}(f)$:

\begin{theorem}
Define $f$ on $N$ variables such that $f(X)=1$ iff $|X|\geq N/10$.
Then $Q^{\unif}(f)$ and $R^{\unif}(f)$ are $O(1)$ and 
$D^{\unif}(f)\in\Omega(N)$.
\end{theorem}

\begin{proof}
Suppose we randomly sample $k$ bits of the input.
Let $a=|X|/N$ denote the fraction of 1s in the input and $\tilde{a}$
the fraction of 1s in the sample.
The Chernoff bound (see e.g.~\cite{alon&spencer:probmethod})
implies that there is a constant $c>0$ such that
$$
\Pr[\tilde{a}<2/10\mid a\geq 3/10]\leq 2^{-ck}.
$$
Now consider the following randomized algorithm for $f$:
\begin{enumerate}
\item
Let $i=100$.
\item
Sample $k_i=i/c$ bits. If the fraction
$\tilde a_i$ of 1s is $\geq 2/10$, then output 1 and stop.
\item
If $i<\log N$, then increase $i$ by 1 and repeat step 2.
\item
If $i\geq\log N$, then count $|X|$ exactly using $N$ queries and
output the correct answer.
\end{enumerate}
It is easy to see that this is a bounded-error algorithm for $f$.
Let us bound its average-case complexity under the uniform distribution.

If $a\geq 3/10$, the expected number of queries for step 2 is 
$$
\sum_{i=100}^{\log N} \Pr[\tilde a_1\leq 2/10, \ldots,
\tilde a_{i-1}\leq 2/10 \mid a\geq 3/10]\cdot\frac{i}{c} \leq 
$$
$$
\sum_{i=100}^{\log N} \Pr[\tilde a_{i-1}\leq 2/10 \mid a \geq3/10]\cdot
\frac{i}{c} \leq \sum_{i=100}^{\log N} 2^{-(i-1)} \cdot\frac{i}{c} \in O(1).
$$
The probability that step 4 is needed (given $a\geq 3/10$)
is at most $2^{-c \log N/c}=1/N$. 
This adds $\frac{1}{N} N=1$ to the expected number of queries.

Under the uniform distribution, the probability of the event  
$a<3/10$ is at most $2^{-c' N}$ for some constant $c'$.
This case contributes at most $2^{-c' N} (N+(\log N)^2) \in o(1)$ to
the expected number of queries. 
Thus in total the algorithm uses $O(1)$ queries on average, 
hence $R^{\unif}(f)\in O(1)$.
Since $Q^{\unif}(f)\leq R^{\unif}(f)$, 
we also have $Q^{\unif}(f)\in O(1)$.

Since a deterministic classical algorithm for $f$ must be correct
on every input $X$, it is easy to see that it must make at least 
$N/10$ queries on every input, hence $D^{\unif}(f)\geq N/10$.
\end{proof}

Accordingly, we can have huge gaps between $D^{\unif}(f)$ and $Q^{\unif}(f)$.
However, this example tells us nothing about the gaps between quantum
and classical bounded-error algorithms.
In the next section we exhibit an $f$ where $Q^{\unif}(f)$ is
exponentially smaller than the classical 
bounded-error complexity $R^{\unif}(f)$.

\section{Exponential Gap between $R^{unif}(f)$ and $Q^{unif}(f)$}\label{secgapunif}

\subsection{The Function}

We use the following modification of Simon's problem~\cite{simon:power}:%
\footnote{The preprint~\cite{hhz:aesuperiority} independently proves a related 
but incomparable result about another Simon-modification.}

\medskip

\noindent
{\bf Input:}
$X=(x_1,\ldots,x_{2^n})$, where each $x_i\in\01^n$.

\noindent
{\bf Output:}
$f(X)=1$ iff there is a non-zero $k\in\01^n$ such that
for all $i\in\01^n$ we have $x_{i\oplus k}=x_i$.

\medskip

Here we treat $i\in\01^n$ both as an $n$-bit string and as a number
between $1$ and $2^n$, and $\oplus$ denotes bitwise XOR.
Note that this function is total (unlike Simon's).
Formally, $f$ is not a Boolean function because the
variables are $\01^n$-valued. However, we can replace every
variable $x_i$ by $n$ Boolean variables and then $f$ becomes a
Boolean function of $N=n 2^n$ variables.
The number of queries needed to compute the Boolean function is at
least the number of queries needed to compute the function with
$\01^n$-valued variables (because we can simulate a query to
the Boolean oracle by means of a query to the $\01^n$-valued input-variables, 
just ignoring the $n-1$ bits that we are not interested in)
and at most $n$ times the number of queries to the $\01^n$-valued oracle
(because one $\01^n$-valued query can be simulated using $n$
Boolean queries). As the numbers of queries are so closely
related, it does not make a big difference whether we use
the $\01^n$-valued oracle or the Boolean oracle. 
For simplicity we count queries to the $\01^n$-valued oracle.

We are interested in the average-case complexity of this function.
The main result is the following exponential gap, to be proven
in the next sections:

\begin{theorem}\label{thuniformgap}
For $f$ as above, 
$Q^{unif}(f)\leq 22n+1$ and $R^{unif}(f)\in\Omega(2^{n/2})$.
\end{theorem}

\subsection{Quantum Upper Bound}

The quantum algorithm is similar to Simon's.
Start with the 2-register superposition 
$\sum_{i\in\01^n}\ket{i}\ket{0}$
(for convenience we ignore normalizing factors).
Apply the oracle once to obtain
$$
\sum_{i\in\01^n}\ket{i}\ket{x_i}.
$$
Measuring the second register gives some $j$ 
and collapses the first register to
$$
\sum_{i:x_i=j}\ket{i}.
$$
A {\em Hadamard transform} $H$ maps bits 
$\ket{b}\rightarrow\frac{1}{\sqrt{2}}(\ket{0}+(-1)^b\ket{1})$.
Applying this to each qubit of the first register gives 
\begin{equation}\label{eqnsupsimon}
\sum_{i:x_i=j}\sum_{i'\in\01^n}(-1)^{(i,i')}\ket{i'}.
\end{equation}
Here $(a,b)$ denotes inner product mod 2;
if $(a,b)=0$ we say $a$ and $b$ are orthogonal.

If $f(X)=1$, then there is a non-zero $k$ such that $x_i=x_{i\oplus k}$ 
for all $i$. In particular, $x_i=j$ iff $x_{i\oplus k}=j$.
Then the final state~(\ref{eqnsupsimon}) can be rewritten as
\begin{eqnarray*}
\sum_{i'\in\01^n}\sum_{i:x_i=j}(-1)^{(i,i')}\ket{i'} & = &
\sum_{i'\in\01^n}\left(\sum_{i:x_i=j}\frac{1}{2}((-1)^{(i,i')}+(-1)^{(i\oplus
k,i')})\right)\ket{i'}\\
 & = & \sum_{i'\in\01^n}\left(\sum_{i:x_i=j}\frac{(-1)^{(i,i')}}{2}(1+(-1)^{(k,i')})\right)\ket{i'}.
\end{eqnarray*}
Notice that $\ket{i'}$ has non-zero amplitude only if $(k,i')=0$.
Hence if $f(X)=1$, then measuring the final state gives some $i'$
orthogonal to the unknown $k$.

To decide if $f(X)=1$, we repeat the above process $m=22 n$ times.
Let $i_1,\ldots,i_m\in\01^n$ be the results of the $m$ measurements.
If $f(X)=1$, there must be a non-zero $k$ that is orthogonal to all $i_r$.
Compute the subspace $S\subseteq\01^n$ that is generated by $i_1,\ldots,i_m$
(i.e.~$S$ is the set of binary vectors obtained by taking linear combinations
of $i_1,\ldots,i_m$ over $GF(2)$).
If $S=\01^n$, then the only $k$ that is orthogonal to all $i_r$ is
$k=0^n$, so then we know that $f(X)=0$.
If $S\neq\01^n$, we just query all $2^n$ values 
$x_{0\ldots 0},\ldots,x_{1\ldots 1}$ and then compute $f(X)$. 
Of course, this latter step is very expensive, but 
it is needed only rarely:

\begin{lemma}
Assume that $X=(x_{0\ldots 0},\ldots,x_{1\ldots 1})$ is chosen
uniformly at random from $\01^N$. 
Then, with probability at least $1-2^{-n}$, $f(X)=0$ and 
the measured $i_1, \ldots, i_m$ generate $\01^n$.
\end{lemma}

\begin{proof}
It can be shown by a small modification 
of~\cite[Theorem~5.1, p.91]{alon&spencer:probmethod}
that with probability at least $1-2^{-c2^{n}}$ ($c>0$),
there are at least $2^n/8$ values $j$ such that
$x_i=j$ for exactly one $i\in\01^n$ (and hence $f(X)=0$).
We assume that this is the case in the following.

If $i_1,\ldots,i_m$ generate a proper subspace of $\01^n$, then
there is a non-zero $k\in\01^n$ that is orthogonal to this subspace.
We estimate the probability that this happens.
Consider some fixed non-zero vector $k\in\01^n$. 
The probability that $i_1$ and $k$ are
orthogonal is at most $\frac{15}{16}$, as follows. 
With probability at least 1/8, the measurement of the second register 
gives $j$ such that $f(i)=j$ for a unique $i$. 
In this case, the measurement of the final superposition~(\ref{eqnsupsimon})
gives a uniformly random $i'$. 
The probability that a uniformly random $i'$ has $(k,i')\neq 0$ is 1/2.
Therefore, the probability that $(k,i_1)=0$ is at most 
$1-\frac{1}{8}\cdot\frac{1}{2}=\frac{15}{16}$.

The vectors $i_1,\ldots,i_m$ are chosen independently.
Therefore, the probability that $k$ is orthogonal to each of them
is at most $(\frac{15}{16})^m=(\frac{15}{16})^{22 n}<2^{-2n}$.
There are $2^n-1$ possible non-zero $k$, so the
probability that there is a $k$ which is orthogonal to each of
$i_1,\ldots,i_m$, is $\leq(2^n-1)2^{-2n}<2^{-n}$.
\end{proof}

Note that this algorithm is actually a {\em zero-error} algorithm:
it always outputs the correct answer.
Its expected number of queries on a uniformly random
input is at most $m=22n$ for generating $i_1, \ldots, i_m$ and at
most $\frac{1}{2^n} 2^n=1$ for querying all the $x_i$ if the first
step does not give $i_1,\ldots,i_m$ that generate $\01^n$.
This completes the proof of the first part of Theorem~\ref{thuniformgap}.
In contrast, in the appendix we show that the {\em worst-case}
zero-error quantum complexity of $f$ is $\Omega(N)$, which is near-maximal.

\subsection{Classical Lower Bound}

Let $D_1$ be the uniform distribution over all inputs $X\in\01^N$ and
$D_2$ be the uniform distribution over all $X$ for which there is a
unique $k\neq 0$ such that $x_i=x_{i\oplus k}$ (and hence $f(X)=1$).
We say an algorithm $A$ {\em distinguishes} between $D_1$ and $D_2$ 
if the average probability that $A$ outputs 0 is $\geq 2/3$ under $D_1$ 
and the average probability that $A$ outputs 1 is $\geq 2/3$ under $D_2$.

\begin{lemma}\label{clb1}
If there is a bounded-error algorithm $A$ that computes $f$ with $m=T_A^{unif}$
queries on average, then there is an algorithm that distinguishes 
between $D_1$ and $D_2$ and uses $O(m)$ queries on all inputs.
\end{lemma}

\begin{proof}
Without loss of generality we assume $A$ has error probability $\leq 1/10$.
To distinguish $D_1$ and $D_2$, we run $A$ until it stops or 
makes $10m$ queries. If it stops, we output the result of $A$.
If it makes $10m$ queries and has not stopped yet, we output 1.

Under $D_1$, the probability that $A$ outputs 1 is 
at most $1/10+o(1)$ ($1/10$ is the maximum probability of error 
on an input with $f(X)=0$ and $o(1)$ is the probability of getting 
an input with $f(X)=1$), so the probability that $A$ outputs 0
is at least $9/10-o(1)$.
The average probability (under $D_1$) that $A$ does not stop 
before $10m$ queries is at most $1/10$, for otherwise 
the average number of queries would be more than $\frac{1}{10}(10m)=m$. 
Therefore the probability under $D_1$ that $A$ outputs 0
after at most $10m$ queries, is at least $(9/10-o(1))-1/10=4/5-o(1)$.
In contrast, the $D_2$-probability that $A$ outputs 0
is $\leq 1/10$ because $f(X)=1$ for any input $X$ from $D_2$.
This shows that we can distinguish $D_1$ from $D_2$.
\end{proof}

\begin{lemma}\label{clb2}
A classical randomized algorithm $A$ that makes $m\in o(2^{n/2})$ queries
cannot distinguish between $D_1$ and $D_2$.
\end{lemma}

\begin{proof}
For a random input from $D_1$,
the probability that all answers to $m$ queries are different is
$$
1\cdot \left(1-\frac{1}{2^n}\right) \cdots \left(1-\frac{(m-1)}{2^n}\right)\geq 1-\sum_{i=1}^{m-1}\frac{i}{2^n}=1-\frac{m(m-1)}{2^{n+1}} = 1-o(1).
$$
For a random input from $D_2$, the probability that there is an $i$ such that 
$A$ queries both $x_i$ and $x_{i\oplus k}$ ($k$ is the hidden vector) 
is $\leq {m\choose 2}/(2^n-1)\in o(1)$, since:
\begin{enumerate}
\item 
for every pair of distinct $i,j$, the probability that
$i=j\oplus k$ is $1/(2^n-1)$
\item
since $A$ queries only $m$ of the $x_i$, it queries only ${m\choose 2}$
distinct pairs $i,j$
\end{enumerate}
If no pair $x_i$, $x_{i\oplus k}$ is queried, 
the probability that all answers are different is
$$
1\cdot \left(1-\frac{1}{2^{n-1}}\right) \cdots \left(1-\frac{(m-1)}{2^{n-1}}\right) = 1-o(1).
$$
It is easy to see that all sequences of $m$ different answers are equally
likely. Therefore, for both distributions $D_1$ and $D_2$, 
we get a uniformly random sequence of $m$ different values with 
probability $1-o(1)$ and something else with probability $o(1)$. 
Thus $A$ cannot ``see'' the difference between $D_1$ and $D_2$ 
with sufficient probability to distinguish between them.
\end{proof}

The second part of Theorem~\ref{thuniformgap} now follows: 
a classical algorithm that computes $f$ with an average number of $m$ queries 
can be used to distinguish between $D_1$ and $D_2$ with $O(m)$ queries 
(Lemma~\ref{clb1}), but then $O(m)\in\Omega(2^{n/2})$ 
(Lemma~\ref{clb2}).

\section{Super-Exponential Gap for Non-Uniform $\mu$}\label{secnonunifgap}

The last section gave an exponential gap between $Q^\mu$
and $R^\mu$ under uniform $\mu$.
Here we show that the gap can be even larger for non-uniform $\mu$.
Consider the average-case complexity of the \OR-function.
It is easy to see that $D^{unif}(\OR)$, $R^{unif}(\OR)$, and $Q^{unif}(\OR)$
are all $O(1)$, since the average input will have many 1s under the 
uniform distribution.
Now we give some examples of non-uniform distributions $\mu$ where 
$Q^\mu(\OR)$ is super-exponentially smaller than $R^\mu(\OR)$:

\begin{theorem}\label{thoravgap}
If $\alpha\in(0,1/2)$ and $\mu(X)=c/{N\choose |X|}(|X|+1)^\alpha(N+1)^{1-\alpha}$ 
($c\approx 1-\alpha$ is a normalizing constant), then
$R^\mu(\OR)\in\Theta(N^\alpha)$ and $Q^\mu(\OR)\in\Theta(1)$.
\end{theorem}

\begin{proof}
Any classical algorithm for \OR\ requires 
$\Theta(N/(|X|+1))$ queries on an input $X$. The upper bound follows from 
random sampling, the lower bound from a block-sensitivity 
argument~\cite{nisan:pram&dt}. 
Hence (omitting the intermediate $\Theta$s):
$$
R^\mu(\OR)=\sum_X\mu(X)\frac{N}{|X|+1}=
\sum_{t=0}^N\frac{cN^\alpha}{(t+1)^{\alpha+1}}
\in\Theta(N^\alpha),
$$
where the last step can be shown by approximating 
the sum over $t$ with an integral.
Similarly, for a quantum algorithm
$\Theta(\sqrt{N/(|X|+1))}$ queries are necessary and 
sufficient on an input $X$~\cite{grover:search,bbht:bounds}, so
$$
Q^\mu(\OR)=\sum_X\mu(X)\sqrt{\frac{N}{|X|+1}}=
\sum_{t=0}^N\frac{cN^{\alpha-1/2}}{(t+1)^{\alpha+1/2}}
\in\Theta(1).
$$
\end{proof}

In particular, for $\alpha=1/2-\eps$ we have the very large 
gap of $O(1)$ quantum versus $\Omega(N^{1/2-\eps})$ classical.
Note that we obtain this super-exponential gap by weighing the 
complexity of two algorithms (classical and quantum \OR-algorithms) 
which are only quadratically apart on each input $X$.
This is the phenomenon we referred to at the end of Section~\ref{secdefs}.

\section{General Bounds for Average-Case Complexity}

In this section we prove some general bounds.
First we make precise the intuitively obvious fact that if an algorithm
$A$ is faster on every input than another algorithm $B$, then it is also 
faster on average under any distribution:

\begin{theorem}
If $\phi:\R\rightarrow\R$ is a concave function and $T_A(X)\leq \phi(T_B(X))$ 
for all $X$, then
$\displaystyle T^\mu_A\leq \phi\left(T^\mu_B\right)$ for every $\mu$.
\end{theorem}

\begin{proof}
By Jensen's inequality, if $\phi$ is concave then 
$E_\mu[\phi(T)]\leq\phi(E_\mu[T])$, hence
$$
T^\mu_A
=\sum_{X\in\01^N}\mu(X)T_A(X)
\leq\sum_{X\in\01^N}\mu(X)\phi(T_B(X))
\leq\phi\left(\sum_{X\in\01^N}\mu(X)T_B(X)\right)=\phi\left(T^\mu_B\right).
$$
\end{proof}

In words: taking the average cannot make the complexity-gap between 
two algorithms smaller. For instance, if $T_A(X)\leq\sqrt{T_B(X)}$
(say, $A$ is Grover's algorithm and $B$ is a classical algorithm for \OR),
then $T^\mu_A\leq\sqrt{T^\mu_B}$.
On the other hand, taking the average {\em can} make the gap much larger, 
as we saw in Theorem~\ref{thoravgap}:
the quantum algorithm for \OR\ runs only quadratically faster than
any classical algorithm on each input, but the {\em average-case} gap 
between quantum and classical can be much bigger than quadratic.

We now prove a general lower bound on $R^\mu$ and $Q^\mu$.
The classical case of the following lemma was shown in~\cite{nisan:pram&dt}, 
the quantum case in~\cite{bbcmw:polynomials}:

\begin{lemma}
Let $A$ be a bounded-error algorithm for some function $f$.
If $A$ is classical then $T_A(X)\in\Omega(bs_X(f))$, and 
if $A$ is quantum then $T_A(X)\in\Omega(\sqrt{bs_X(f)})$.
\end{lemma}

A lower bound in terms of the $\mu$-expected
block sensitivity follows:

\begin{theorem}\label{thavgbsbound}
For all $f$, $\mu$:
$R^\mu(f)\in\Omega(E_\mu[bs_X(f)])$ and 
$Q^\mu(f)\in\Omega(E_\mu[\sqrt{bs_X(f)}])$.
\end{theorem}

\section{Average-Case Complexity of MAJORITY}

Here we examine the average-case complexity of the MAJORITY-function.
The hard inputs for majority occur when $t=|X|\approx N/2$. 
Any quantum algorithm needs $\Omega(N)$ queries for such 
inputs~\cite{bbcmw:polynomials}.   
Since the uniform distribution puts most probability on the set of 
$X$ with $|X|$ close to $N/2$, we might expect an $\Omega(N)$ average-case
complexity as well. 
However, we will prove that the complexity is nearly $\sqrt{N}$.
For this we need the following result about approximate quantum counting, 
which is Theorem~13 of~\cite{bhmt:countingj} 
(this is the upcoming journal version of~\cite{bht:counting} 
and~\cite{mosca:eigen}; see also~\cite[Theorem~1.10]{nayak&wu:median}):

\begin{theorem}[Brassard, H{\o}yer, Mosca, Tapp]\label{thqcounting}
There exists a quantum algorithm {\bf QCount} with the following property.
For every $N$-bit input $X$ (with $t=|X|$) and number of queries $T$,
and any integer $k\geq 1$, {\bf QCount} uses $T$ queries and
outputs a number $\tilde{t}$ such that
$$
|t-\tilde{t}|\leq 2\pi k\frac{\sqrt{t(N-t)}}{T}+\pi^2k^2\frac{N}{T^2}
$$
with probability at least $8/\pi^2$ if $k=1$ and
probability $\geq 1- 1/2(k-1)$ if $k>1$.
\end{theorem}

Using repeated applications of this quantum counting routine
we can obtain a quantum algorithm for majority that is fast on average:

\begin{theorem}\label{thavqmaj}
$Q^{\unif}(\MAJ)\in O(\sqrt{N}(\log N)^2)$.
\end{theorem}

\begin{proof}
For all $i\in\{1,\ldots,\log N\}$, 
define $A_i=\{X\mid N/2^{i+1}< \left| |X|-N/2 \right|\leq N/2^i\}$.
The probability under the uniform distribution of getting an input $X\in A_i$
is $\mu(A_i)\in O(\sqrt{N}/2^i)$, since 
the number of inputs $X$ with $k$ 1s is
${N\choose k}\in O(2^N/\sqrt{N})$
for all $k$. The idea of our algorithm is to have $\log N$ runs of
the quantum counting algorithm, with increasing numbers of queries,
such that the majority value of inputs from $A_i$ is probably
detected around the $i$th counting stage. 
We will use $T_i=100\cdot 2^i\log N$ queries in the $i$th counting stage.
Our MAJORITY-algorithm is the following:
\begin{quote}
For $i=1$ to $\log N$ do:\\[1mm]
\hspace*{7mm}quantum count $|X|$ using $T_i$ queries 
(call the estimate $\tilde{t}_i$)\\
\hspace*{7mm}if $|\widetilde{t}_i-N/2|>N/2^i$, then output whether
$\widetilde{t}_i>N/2$ and stop.\\[1mm]
Classically count $|X|$ using $N$ queries and output its majority.
\end{quote}
Let us analyze the behavior of the algorithm on an input $X\in A_i$.
For $t=|X|$, we have $|t-N/2|\in(N/2^{i+1},N/2^i]$. 
By Theorem~\ref{thqcounting}, with probability $>1-1/10\log N$
we have $\left|\widetilde{t}_i-t\right|\leq N/2^i$, so
with probability $(1-1/10\log N)^{\log N}\approx e^{-1/10}>0.9$ we have
$\left|\widetilde{t}_i-t\right|\leq N/2^i$ for all $1\leq i\leq N$.
This ensures that the algorithm outputs the correct value 
with high probability. 

We now bound the expected number of queries the algorithm needs on input $X$.
Consider the $(i+2)$nd counting stage. 
With probability $1-1/10\log N$ we will have
$|\tilde{t}_{i+2}-t|\leq N/2^{i+2}$. 
In this case the algorithm will terminate, because
$$
|\tilde{t}_{i+2}-N/2|\geq 
|t-N/2|-|\tilde{t}_{i+2}-t|>N/2^{i+1}-N/2^{i+2}=N/2^{i+2}.
$$
Thus with high probability the algorithm needs no more than $i+2$ counting 
stages on input $X$. Later counting stages take exponentially 
more queries ($T_{i+2+j}=2^jT_{i+2}$), but are needed only with 
exponentially decreasing probability $O(1/2^j\log N)$:
the probability that $|\tilde{t}_{i+2+j}-t|>N/2^{i+2}$ goes down exponentially 
with $j$ precisely because the number of queries goes up exponentially.
Similarly, the last step of the algorithm (classical counting) is needed only 
with negligible probability. 

Now the expected number of queries on input $X$ can be upper bounded by
$$
\sum_{j=1}^{i+2}T_i + \sum_{k=i+3}^{\log N}T_k\cdot 
O\left(\frac{1}{2^{k-i-3}\log N}\right)
< 100\cdot 2^{i+3}\log N + \sum_{k=i+3}^{\log N}100\cdot 2^{i+3}
\in O(2^i\log N).
$$
Therefore under the uniform distribution the average
expected number of queries can be upper bounded by
$\sum_{i=1}^{\log N} \mu(A_i)O(2^i\log N)\in O(\sqrt{N}(\log N)^2).$
\end{proof}

The nearly matching lower bound is:

\begin{theorem}\label{thavqmajlower}
$Q^{\unif}(\MAJ)\in\Omega(\sqrt{N})$.
\end{theorem}

\begin{proof}
Let $A$ be a bounded-error quantum algorithm for MAJORITY.
It follows from the worst-case results of~\cite{bbcmw:polynomials}
that $A$ uses $\Omega(N)$ queries on the hardest inputs, which
are the $X$ with $|X|=N/2\pm 1$.
Since the uniform distribution puts $\Omega(1/\sqrt{N})$ probability on
the set of such $X$, the average-case complexity of $A$ is at least
$\Omega(1/\sqrt{N})\Omega(N)=\Omega(\sqrt{N})$.
\end{proof}

What about the {\em classical} average-case complexity of MAJORITY?
Alonso, Reingold, and Schott~\cite{ars:avmaj} 
prove the bound $D^{\unif}(\MAJ)=2N/3-\sqrt{8N/9\pi}+O(\log N)$ 
for deterministic classical computers.
We can also prove a linear lower bound for the {\em bounded-error} 
classical complexity, using the following lemma:

\begin{lemma}
Let $\Delta\in\{1,\ldots,\sqrt{N}\}$.
Any classical bounded-error algorithm that computes MAJORITY
on inputs $X$ with $|X|\in\{N/2,N/2+\Delta\}$ must make $\Omega(N)$ 
queries on all such inputs.
\end{lemma}

\begin{proof}
We will prove the lemma for $\Delta=\sqrt{N}$, which is the hardest case.
We assume without loss of generality that the algorithm queries 
its input $X$ at $T(X)$ random positions, and outputs 1 if the fraction 
of 1s in its sample is at least $(N/2+\Delta)/N=1/2+1/\sqrt{N}$.
We do not care what the algorithm outputs otherwise.
Consider an input $X$ with $|X|=N/2$. The algorithm uses $T=T(X)$
queries and should output 0 with probability at least $2/3$.
Thus the probability of output 1 on $X$ must be at most $1/3$, in particular
$$
\Pr[\mbox{ at least $T(1/2+1/\sqrt{N})$ 1s in sample of size $T$}]\leq 1/3.
$$
Since the $T$ queries of the algorithm can be viewed as
sampling without replacement from a set containing $N/2$ 1s and $N/2$ 0s, 
this error probability is given by the hypergeometric distribution
$$
\Pr[\mbox{ at least $T(1/2+1/\sqrt{N})$ 1s in sample of size $T$}]=
\frac{\displaystyle \sum_{i=T(1/2+1/\sqrt{N})}^{T}{{N/2}\choose i}\cdot {{N/2}\choose{T-i}}}
{\displaystyle {N\choose T}}.
$$
We can approximate the hypergeometric distribution using the
normal distribution, see e.g.~\cite{nicholson:hyper}.
Let $z_k=(2k-T)/\sqrt{T}$ and 
$\Phi(z)=\int_{-\infty}^z\frac{1}{\sqrt{2\pi}}e^{-t^2/2}dt$,
then the above probability approaches
$$
\Phi(z_T)-\Phi(z_{T(1/2+1/\sqrt{N})}).
$$
Note that $\Phi(z_T)=\Phi(\sqrt{T})\rightarrow 1$ and that
$\Phi(z_{T(1/2+1/\sqrt{N})})=\Phi(2\sqrt{T/N})\rightarrow 1/2$ if $T\in o(N)$.
Thus we can only avoid having an error probability close to 1/2 by using
$T\in\Omega(N)$ queries on $X$ with $|X|=N/2$.
A similar argument shows that we must also use $\Omega(N)$ queries
if $|X|=N/2+\Delta$.
\end{proof}

It now follows that:

\begin{theorem}
$R^{\unif}(\MAJ)\in\Omega(N)$.
\end{theorem}

\begin{proof}
The previous lemma shows that any algorithm for MAJORITY needs
$\Omega(N)$ queries on inputs $X$ with $|X|\in[N/2,N/2+\sqrt{N}]$.
Since the uniform distribution puts $\Omega(1)$ probability on the 
set of such $X$, the theorem follows.
\end{proof}

Accordingly, {\em on average} a quantum computer can compute MAJORITY 
almost quadratically faster than a classical computer, whereas for the
{\em worst-case} input quantum and classical computers are about equally 
fast (or slow).

\section{Average-Case Complexity of PARITY}

Finally we prove some results for the average-case complexity of PARITY. 
This is in many ways the hardest Boolean function.
Firstly, $bs_X(f)=N$ for all $X$, hence by Theorem~\ref{thavgbsbound}:

\begin{corollary}
For every $\mu$,
$R^\mu(\PARITY)\in\Omega(N)$ and $Q^\mu(\PARITY)\in\Omega(\sqrt{N})$.
\end{corollary}

With high probability we can obtain an exact count of $|X|$,
using $O(\sqrt{(|X|+1)N})$ quantum queries~\cite{bhmt:countingj}.
Combining this with a $\mu$ that puts $O(1/\sqrt{N})$ probability 
on the set of all $X$ with $|X|>1$ and distributes the remaining probability
arbitrarily over the $X$ with $|X|\leq 1$, we obtain a distribution
$\mu$ such that $Q^\mu(\PARITY)\in O(\sqrt{N})$. 

We can prove $Q^\mu(\PARITY)\leq N/6$ for any $\mu$
by the following algorithm: with probability $1/3$ output 1, 
with probability $1/3$ output 0, and with probability $1/3$ run 
the exact quantum algorithm for PARITY, which has worst-case 
complexity $N/2$~\cite{bbcmw:polynomials,fggs:parity}.
This algorithm has success probability $2/3$ on every input and has 
expected number of queries equal to $N/6$.

More than a linear speed-up on average is not possible if $\mu$ is uniform:

\begin{theorem}
$Q^{\unif}(\PARITY)\in\Omega(N)$.
\end{theorem}

\begin{proof}
Let $A$ be a bounded-error quantum algorithm for PARITY.
Let $B$ be an algorithm that flips each bit of its input $X$ with
probability $1/2$, records the number $b$ of actual bitflips, runs
$A$ on the changed input $Y$, and outputs $A(Y)+ b\mbox{ mod }2$.
It is easy to see that $B$ is a bounded-error algorithm for PARITY and
that it uses an {\em expected} number of $T_A^\mu$ queries on {\em every} input.
Using standard techniques, we can turn this into
an algorithm for PARITY with {\em worst-case} $O(T_A^\mu)$ queries.
Since the worst-case lower bound for PARITY is 
$N/2$~\cite{bbcmw:polynomials,fggs:parity}, the theorem follows.
\end{proof}

\subsection*{Acknowledgments}
We thank Harry Buhrman for suggesting this topic, and him, Lance Fortnow, 
Lane Hemaspaandra, Hein R\"ohrig, Alain Tapp, and Umesh Vazirani 
for helpful discussions.
Also thanks to Alain for sending a draft of~\cite{bhmt:countingj}.


\appendix
\section{Worst-case Complexity of $f$}

In this appendix we will show a lower bound of $\Omega(N)$ queries
for the zero-error worst-case complexity $Q_0(f)$  
of the function $f$ on $N=n2^n$ binary variables defined 
in Section~\ref{secgapunif}. (We count binary queries this time.)
Consider a quantum algorithm that makes at most $T$ queries 
and that, for every $X$, outputs either the correct output $f(X)$ or,
with probability $\leq 1/2$, outputs ``inconclusive''.
We use the following lemma from~\cite{bbcmw:polynomials}:

\begin{lemma}
The probability that a $T$-query quantum algorithm outputs 1 
can be written as a multilinear $N$-variate polynomial $P(X)$
of degree at most $2T$.
\end{lemma}

Consider the polynomial $P$ induced by our $T$-query algorithm for $f$.
It has the following properties:
\begin{enumerate}
\item $P$ has degree $d\leq 2T$
\item if $f(X)=0$ then $P(X)=0$
\item if $f(X)=1$ then $P(X)\in[1/2,1]$
\end{enumerate}
We first show that only very few inputs $X\in\01^N$ make $f(X)=1$.
The number of such 1-inputs for $f$ is the number of ways to choose 
$k\in\01^n-\{0^n\}$, times the number of ways to choose 
$2^n/2$ independent $x_i\in\01^n$,
which is $(2^n-1)\cdot (2^n)^{2^n/2}<2^{n(2^n/2+1)}$.
Accordingly, the fraction of 1-inputs among all $2^N$ inputs $X$ is
$<2^{n(2^n/2+1)}/2^{n2^n}=2^{-n(2^n/2-1)}$.
These $X$ are exactly the $X$ that make $P(X)\neq 0$.
However, the following result is 
known~\cite{schwartz:probabilistic,nisan&szegedy:degree}:

\begin{lemma}[Schwartz]
If $P$ is a non-constant $N$-variate multilinear polynomial of degree $d$, then 
$$
\frac{|\{X\in\01^N \mid P(X)\neq 0\}|}{2^N}\geq 2^{-d}.
$$
\end{lemma}

This implies $d\geq n(2^n/2-1)$ and hence $T\geq d/2\geq n(2^n/4-2)\approx N/4$.
Thus we have proved that the worst-case zero-error quantum complexity of $f$
is near-maximal:

\begin{theorem}
$Q_0(f)\in\Omega(N)$.
\end{theorem}

\end{document}